\documentclass[aps,manuscript]{revtex4-2} 
\usepackage{bm}
\usepackage{amsmath}
\usepackage{color}
\usepackage{graphicx}
\usepackage{amssymb}
\usepackage[toc,page]{appendix}

\DeclareMathAlphabet{\mathbbmsl}{U}{bbm}{m}{sl}
\makeatletter
\newsavebox{\@brx}
\newcommand{\llangle}[1][]{\savebox{\@brx}{\(\m@th{#1\langle}\)}%
	\mathopen{\copy\@brx\kern-0.5\wd\@brx\usebox{\@brx}}}
\newcommand{\rrangle}[1][]{\savebox{\@brx}{\(\m@th{#1\rangle}\)}%
	\mathclose{\copy\@brx\kern-0.5\wd\@brx\usebox{\@brx}}}
\makeatother
\begin{document}
\draft

\title{Strain Engineering of Magneto-optical Properties in C$_3$B/C$_3$N van der Waals Heterostructure}

\author{Po-Hsin Shih$^{1}$, Thi-Nga Do$^{2,}\footnote{Corresponding author: {\em E-mail}: sofia90vn@gmail.com}$, Godfrey Gumbs$^{1,}\footnote{Corresponding author: {\em E-mail}: ggumbs@hunter.cuny.edu}$}
\affiliation{$^{1}$ Department of Physics and Astronomy, Hunter College of the City University of New York,
695 Park Avenue, New York, New York 10065, USA \\
$^{2}$ Department of Physics, National Cheng Kung University, Tainan 701, Taiwan
}

\date{\today}

\begin{abstract}

Carbon-based bilayer van der Waals (vdW) materials are attracting much attention due to their predicted interesting physical properties. Here, we theoretically investigate electronic and optical properties of C$_3$B/C$_3$N vdW heterostructure (HTS) under external magnetic field and mechanical strain. The tight-binding model of the system is constructed to include the strain-induced modification of the hopping interactions. The influence of a uniform perpendicular magnetic field is included by using the Peierls substitution method. We observe the intriguing electronic and optical characteristics of the HTS under mechanical strain, covering the band inversion, alteration of band gap and optical gap, distortion of band-edge states, as well as significant enhancement of optical absorption. Furthermore, the interplay between external magnetic field and biaxial strain leads to exotic features of quantization and optical spectra. This work provides important information for the comprehension of the engineering of materials by external effects. Our study suggests that C$_3$B/C$_3$N vdW HTS is a promising candidate for next-generation electronic and optoelectronic devices.

\end{abstract}
\maketitle

\section{Introduction}
\label{sec1}

The van der Waals (vdW) heterostructures (HTS) formed by stacking different two-dimensional (2D) materials have been demonstrated to exhibit the enhanced performance compared with the isolated individual monolayers. These materials have attracted much interest due to their fascinating electronic and optical properties. The explored vdW HTS include graphene-based materials \cite{grapheneHTS}, transition-metal dichalcogenide (TMDC) \cite{TMDC-nc, TMDC-nn}, and other compounds \,\cite{compound}. Although the interlayer electronic coupling is rather weak, its critical role in determining the fundamental characteristics of the HTS is indisputable. Understanding how to tune the physical properties of materials by controlling interlayer coupling is desirable. The carbon-based C$_3$B/C$_3$N vdW HTS is a typical p-n junction, composed of two single-layer p-type C$_3$B and n-type C$_3$N, with relatively strong interlayer interaction \cite{c3bc3n-jpcl, c3bc3n-pss}. Both C$_3$B and C$_3$N monolayers possess graphene-like lattice structure.  They have been synthesized by various experimental techniques such as epitaxial method \cite{c3bc3n-prl, c3bc3n-solid, c3bc3n-prb, c3bc3n-jpcm}, polymerization \cite{c3bc3n-adv}, and pyrolysis of organic single crystals \cite{c3bc3n-pnas}. However, these individual materials exhibit large indirect band gaps, which limit them from having  many important applications. On the other hand, the C$_3$B and C$_3$N monolayers have the same lattice parameters so that they can be simply stacked together to form an HTS. Therefore, C$_3$B/C$_3$N vdW HTS is expected to be an ideal system to test the interlayer coupling which is needed for the study of tuning physical properties of the material.

\medskip
\par
Mechanical strain is an efficient way to adjust the atomic interactions of 2D materials, leading to the strain engineering of fundamental physical properties. Strain can be induced and tuned naturally or intentionally on 2D materials by different methods, such as epitaxial growth, mechanical exfoliation, chemical vapor deposition, chemical doping and absorption, intrinsic ripples, thermal expansion mismatch, strain relaxation and critical thickness in HTS, substrate modification, pressurized blisters, blown-bubble, tip indentation, polymer encapsulation, and local strain application \cite{strain-nano, strainTMDC-am}. Note that strain is unavoidable for the designed vdW HTS. For strained 2D materials, the strain distribution can be monitored by various techniques, including scanning tunneling microscopy, high-resolution transmission electron microscopy, Raman and  photoluminescence (PL) spectroscopy, and second and third harmonic generation. So far, many experimental and theoretical studies have shown the significant effect of strain on electronic and optical properties of graphene \cite{strainG-prb, strainG-lsa}, 2D TMDCs \cite{strainTMDC-am, strainTMDC-info}, other 2D HTS \cite{strainHTS-mtc, strainHTS-nano}, and compound systems \cite{straincomp-E, straincomp-nrl}. Considerable findings include band-structure tuning, metal-semiconductor phase transition, inducing optical gap, and enhancement of optical conductivity. Based on the strain-induced modification of optical and electronic properties, many follow-up research works on optoelectronic devices have been performed \cite{device}. Strained 2D materials are potential candidates for next-generation electronic and optoelectronic devices such as flexible strain sensors and photodetectors. The combination of strain and other effects, such as doping, designing vdW HTS, and magnetic field, is expected to engender novel electronic and optical properties of materials. C$_3$B/C$_3$N vdW HTS is an idea platform for the investigation of strain-enriched fundamental properties. This can play a critical role in understanding the electronic and optical properties of 2D materials and their device applications.

\medskip
\par
A magnetic field can diversify considerably the electronic and optical properties of C$_3$B/C$_3$N vdW HTS due to the quantization of electronic states.
The Landau level (LL) energies can be identified by scanning tunneling spectroscopy, quantum Hall transport measurements, and cyclotron resonance as done for graphene and many other 2D materials \cite{LLqhe-nature, LLsts-solid, LLsts-prl, LLcyclo}.
Moreover, their low-energy magneto-optical spectra with many delta-function-like pronounced peaks have been examined by using optical spectroscopies \cite{LLoptical-prl07, LLoptical-prl08}.
The strained-induced pseudo-magnetic field has been predicted theoretically and examined experimentally in graphene and TMDC.
However, it was proved that the pseudo-LL quantization only occurs under specific conditions, such as inhomogeneous strain and lattice distortion \cite{pseudoB-np, pseudoB-prb10, pseudoB-prb15}.
On the other hand, the interplay between an external magnetic field and biaxial strain is expected to trigger interesting electronic and optical properties of 2D materials, which is worthy of a thorough investigation.
For C$_3$B/C$_3$N vdW HTS, doping with either B or N atoms on each layer makes the orbital contribution much more complex than that of bilayer graphene, which enriches substantially the LL features.
The optical transition between such LLs leads to peculiar magneto-optical characteristics.

\medskip
\par
The rest of the paper is organized as follows. In Sec.\,\ref{sec2}, we present our theoretical model for executing numerical computations, the tight-binding model (TBM) combined with Peierls substitution. In Sec.\,\ref{sec3}, we present a detailed discussion on obtained numerical results for the 2D C$_3$B/C$_3$N vdW HTS under strain and external magnetic field. Finally, conclusions are summarized in Sec.\,\ref{sec4}.

\section{Theoretical Method}
\label{sec2}

We construct a TBM for C$_3$B/C$_3$N vdW HTS with two individual layers, forming Bernal stacking configuration, as illustrated in Figs. 1a-1b. When the system is subjected to biaxial strain, either tensile or compressive, the bond length varies proportionally in both x and y directions. Both the intralayer and interlayer atomic interactions will change accordingly. TBM is widely used to investigate the essential physical properties of condensed-matter materials based on the band structure calculation. This method has been successfully applied to a wide range of materials such as graphene and related systems \cite{tbm1, tbm2}, topological materials \cite{tbm3}, HTS \cite{tbm4} and so on. TBM can be combined with other theories to explore specific physical properties, including magnetic quantization \cite{tbm5}, quantum Hall effect \cite{tbm5, tbm6}, optical properties \cite{tbm7}, and electronic excitation \cite{tbm8}. Several theoretical approaches have been commonly used to study similar physical properties, such as the continuum model and the${\bf  k}\cdot{\bf p}$ perturbation.
However, those theoretical methods consider separate models for each stress without any connection whereas our TBM shows a strong relationship between lattice structures, atomic orbitals and continuously applied strain which allows for a deep understanding of the controlling physical properties by strain effect. We develop a modified TBM in order to take into account the continuously modified strain in terms of hopping parameters and bonding length. This is important for the investigation of engineering physical properties of materials by external effects, leading to possible technology and device applications.

\medskip
\par
In order to reproduce the exact bands as calculated by the first-principles method, we fit the strain-dependent hopping parameters and other coefficients such as site energies and layer potentials. Our tight-binding band structures are in good agreement with the first-principles ones presented in \cite{c3bc3n-jpcl, c3bc3n-pss}, regarding both energy and orbital distributions.
The tight-binding Hamiltonian for C$_3$B/C$_3$N vdW HTS under strain can be written as

\begin{equation}
H  = \sum\limits_{\alpha,\beta;m} \sum\limits_{\ell,\ell^{\prime} = 1}^{2} [t_{\alpha \beta}^{m}(\mathbf{d}_s)(c_{\alpha}^{\ell})^\dagger c_{\beta}^{\ell^{\prime}} + \delta_{\ell \ell^{\prime}}\nu_{\ell} + \delta_{\ell \ell^{\prime}} \delta_{\alpha \beta} (\epsilon_{\alpha} + \epsilon (\mathbf{d}_s) )  ] + H.c.\ .
\end{equation}
In this notation, $\alpha$ and $\beta$ denote the $p_z$ orbitals of atoms and $\ell$ and $\ell^{\prime}$ stand for the HTS layers. $t_{\alpha \beta}^{m}(\mathbf{d}_s)$ is the strain-dependent hopping term, in which $\mathbf{d}_s = (1+s)d_0$ represents the strain-dependent vector connecting two atoms while $s \in [-3; 3] \% $ is our considered strain strength range and $d_0$ = 1.42 $\AA$ is unstrained bond length; $m = 0, 1, 2$ specifies, respectively, the hopping interaction between two intralayer nearest, interlayer nearest (vertical) and interlayer next-nearest atoms.
The term $\epsilon(\mathbf{d}_s)= s \epsilon_0$ ($\epsilon_0 = 0.0226$ eV) defines the strain-dependent on-site energy.
The annihilation operator $c_{\alpha}^{\ell}$ (creation operator $(c_{\alpha}^{\ell})^{\dagger}$) can destroy (generate) an $\alpha$ electronic state on the $\ell$-th layer. $\nu_{1} = 0.0929$ eV and $\nu_{1} = 0.0606$ eV are, respectively, the C$_3$B and C$_3$N layer potentials. The site energies of three atoms are estimated as $\epsilon_C = 0$, $\epsilon_B = 3.3696$ eV, and $\epsilon_N = -3.7953$ eV. In addition, H.c. indicates the Hermitian conjugate.

\medskip
\par

\begin{figure}[htbp]
\begin{center}
\includegraphics[width=0.9\linewidth]{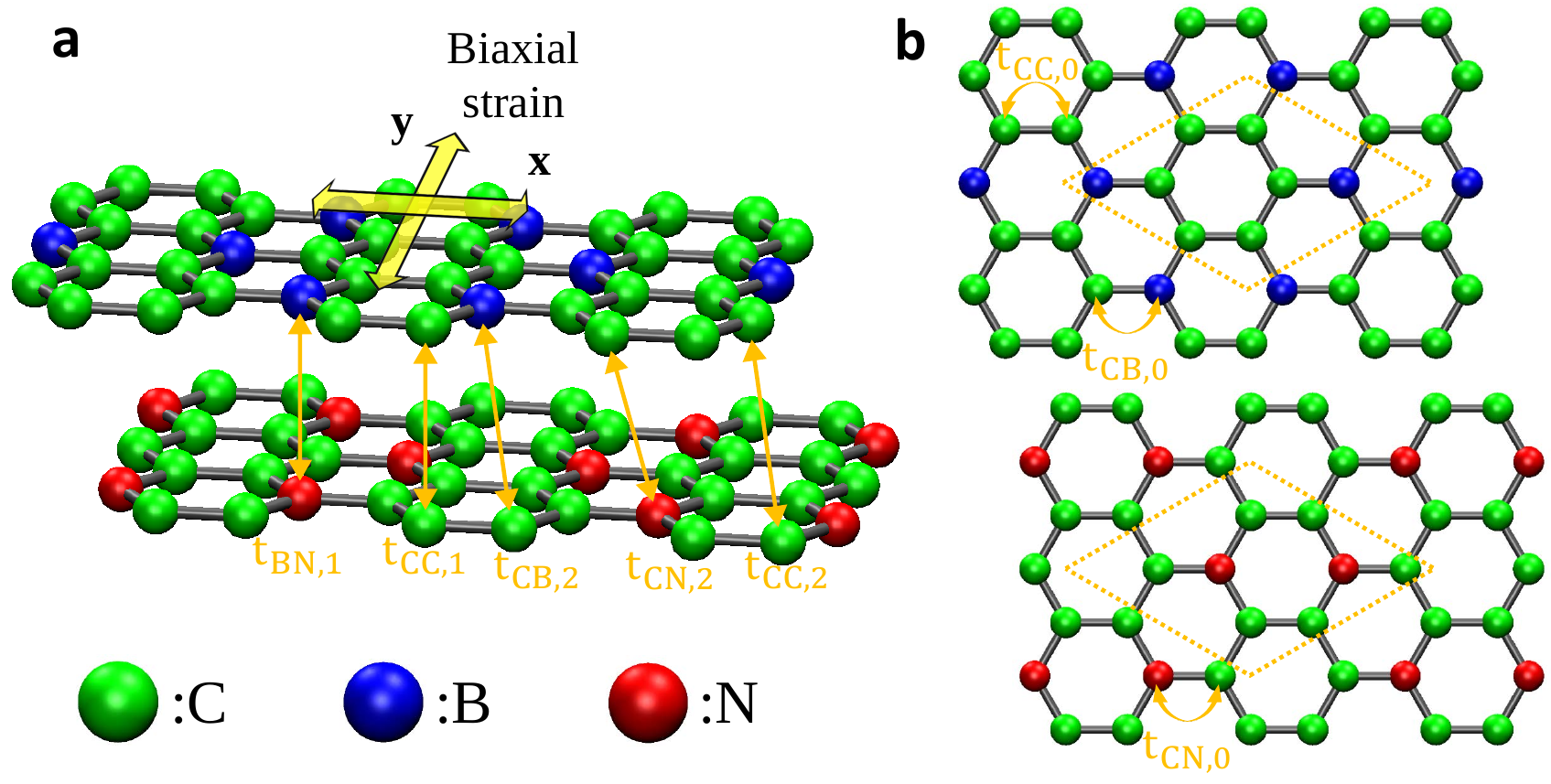}
\end{center}
\caption{(color online) Lattice structure of C$_3$B/C$_3$N vdW HTS. Part (a) illustrates the side view, in which both the upper C$_3$B and lower C$_3$N layers are honeycomb crystals, stacking together to form Bernal stacking configurations. The orange arrows show the intralayer and interlayer hopping interactions. The two crossing yellow arrows indicate the directions of the applied biaxial strain. Part (b) displays the top views of two individual layers. Three types of intralayer hopping terms are marked by the orange arrows. The orange rhombus shapes mark the unit cell projections onto the two layers. The unit cell of C$_3$B/C$_3$N vdW HTS consists of 16 atoms. }
\label{Fig1}
\end{figure}

The hopping interaction terms $t_{\alpha \beta}^{m}(\mathbf{d}_s)$ are optimized by employing the frequently used Slater-Koster two-center approximation \cite{Harrison}. To consider continuous strain, we develop a formula to fit the strain-dependent intralayer and interlayer hopping parameters as follows.

\begin{equation}
t_{\alpha \beta}^{m}(\mathbf{d}_s) = [ 1- \frac{\mathbf{d}_z^2}{d_s^2} ] \frac{V_{\pi \alpha \beta}^{m}}{d_s^2} e^{-\frac{|\mathbf{d}_s|-b}{\delta_1}} + \frac{\mathbf{d}_z^2}{d_s^2} \frac{V_{\sigma \alpha \beta}^{m}}{d_s^2} e^{-\frac{|\mathbf{d}_s|-b}{\delta_2}},
\end{equation}
where $d_s$ is the magnitude and $\mathbf{d}_z$ is the z-component of $\mathbf{d}_s$, $b = 1.42 \AA$ is the distance between two nearest atoms, $\pi$ and $\sigma$ label the bands, and $\delta_1 = -3.0498$ and $\delta_2 = 9.7967$ are chosen as the decay lengths of the transfer integral. The exponential terms assure that the transfer integral for large $d$ is sufficiently small and can be reasonably neglected. Furthermore, the coefficients $V_{\pi \alpha \beta}^{m}$ and $V_{\sigma \alpha \beta}^{m}$ can be found in Table 1.
\medskip
\par

\begin{table}[h]
  \begin{center}
    \caption{The coefficients $V_{\pi \alpha \beta}^{m}$ and $V_{\sigma \alpha \beta}^{m}$ of the hopping integrals. The first column shows the notation of coefficients for different orbitals and $m$'s. The values of intralayer and interlayer atomic interactions among C, B, and N are given in the last four columns. Note that, C-B means $\alpha$ = C and $\beta$ = B.}
    \label{tab:table1}
    \begin{tabular}{|c|c|c|c|c|}
      \hline
      \text{Coefficients} & \text{C-C} & \text{C-B} & \text{C-N} & \text{B-N}\\
      \hline
      $V_{\pi \alpha \beta}^{0}$ & 22.8933 & 7.9012 & 8.1499 & 0  \\
      $V_{\pi \alpha \beta}^{1}$ & 0.0204 & 0 & 0 & -0.0017  \\
      $V_{\sigma \alpha \beta}^{1}$ & -0.0363 & 0 & 0 & 0.0762  \\
      $V_{\pi \alpha \beta}^{2}$ & -10.9659 & -19.6753 & -0.2861 & 0  \\
      $V_{\sigma \alpha \beta}^{2}$ & -9.7118 & -7.3308 & 0.8867 & 0  \\
      \hline
    \end{tabular}
  \end{center}
\end{table}
The effect of an external magnetic field is taken into consideration by using the Peierls substitution. A uniform perpendicular magnetic field $\mbox{\boldmath$B$}= (0,0,B_0)$ can modify significantly the lattice periodicity, giving rise to the change in the tight-binding Hamiltonian. Such influence of external magnetic field on the system can be taken into account by employing the Peierls substitution, in which the Peierls phase takes the form $G_{R} = (2\pi/\phi_{0})\int\limits_{\bf R}^{\bf r} \mbox{\boldmath$A$}\cdot d\mbox{\boldmath$\ell$}$ \cite{tbm5, PsubPei}. In this notation, $\mbox{\boldmath$A$}=(0,B_0x,0)$ is the magnetic vector potential within the Landau gauge and $\phi_{0}=hc/e$ is the flux quantum. The Peierls phases are added into the hopping terms to imply the magnetic-field-induced expansion of the unit cell along the x direction. The number of atoms in a magnetic unit cell, which strongly depends on the strength of the applied field, is defined as $2N \times 2\phi_{0}/ \phi$ with $N = 16$ and $2\phi_{0}/ \phi$ being the number of atoms in a primitive unit cell and a period of the Peierls phase, respectively. Here, the magnetic flux $\phi$ is determined as $\phi = B_0{\cal S}$ where ${\cal S}$ is the area of a magnetic unit cell.

\medskip
\par
When C$_3$B/C$_3$N vdW HTS is subject to an electromagnetic field carrying the electric polarization $\hat{\mathbf{E}}$ with frequency $\omega$, the vertical optical transitions from occupied valence ($v$) to unoccupied conduction ($c$) states will be triggered. The absorption function for vertical transition can be expressed as \cite{abfunction}

\begin{eqnarray}
A(\omega) \propto
&\sum_{c,v,i,j} \int_{1stBZ} \frac {d\mathbf{k}}{(2\pi)^2}
 \Big| \Big\langle \Psi^{c} (\mathbf{k},j)
 \Big| \frac{   \hat{\mathbf{E}}\cdot \mathbf{P}   } {m_e}
 \Big| \Psi^{v}(\mathbf{k},i)    \Big\rangle \Big|^2 \nonumber \\
 &\times
Im \Big[      \frac{f(E^c (\mathbf{k},j)) - f(E^v (\mathbf{k},i))}
{E^c (\mathbf{k},j)-E^v (\mathbf{k},i)-\omega - i\Gamma}           \Big].
\end{eqnarray}
In this notation, $\mathbf{P}$ is the momentum operator, $f(E^{c,v}(k,i))$ represents the Fermi-Dirac distribution function, $m_e$ stands for the free-electron mass, and ($i$, $j$) denote the lattice sites. The broadening factor, $\Gamma$, is chosen to be sufficiently small (= 1 meV) in our calculations. The absorption transition corresponding to the non-vanishing velocity matrix element, $\Big\langle \Psi^{c} (\mathbf{k},j)  \Big| \frac{   \hat{\mathbf{E}}\cdot \mathbf{P}   } {m_e}  \Big| \Psi^{v}(\mathbf{k},i)    \Big\rangle$, will be revealed in the optical spectrum. This term can be estimated by using the gradient approximation \cite{gradient}.

\medskip
\par
The eigenvalues and wave functions of the Hamiltonian are critical for the calculation of the absorption function to investigate the optical properties of materials. Within the tight-binding approximation, the wave function is defined as a superposition of the product of the coefficient and the 2$p_z$ atomic orbital function. The eigenvalues of the Hamiltonian and the coefficient of the tight-binding function can be obtained directly from solving the Hamiltonian matrix. Note that the orbital functions are orthonormal, thus the relevant inner product term in the velocity matrix element become a delta function. The tight-binding functions are then inserted into the velocity matrix element term of the absorption function above to compute the absorption coefficients. It is worth mentioning that, the significant effect due to phonon on essential physical properties of condensed matter systems have been reported previously \cite{phonon-npj, phonon-prb}. Phonons are generated by lattice vibration, which might arise from atoms own thermal energy or outside forces such as temperature and lattice defect \cite{phonon-npj, phonon-prb}. Here we consider free-standing HTS at zero temperature under biaxial strain without lattice distortion. Therefore, the phonon effect is neglected in our calculations.

\medskip
\par

\section{Results and Discussion}
\label{sec3}

The band structures of unstrained and specific strained C$_3$B/C$_3$N vdW HTS are presented in Figs. 2a-2b.
Here we only show the low-lying tight-binding bands nearest to the neutral point because those bands are already located in quite a wide energy range E $\in$ [-3, 3] eV. Moreover, our considered regime is suitable for the focus of this work which is the controlling energy bands and optical spectra by mechanical strain. Note that strain only has significant effect on the low-lying bands, specifically E $\in$ [-0.5, 0.5] eV, as demonstrated in Fig. 2b. These low-lying bands are associated with the absorption spectrum up to $\omega$ = 1 eV or THz regime, which can be probed by infrared spectroscopies. The optical properties within such frequency range have attracted great attention and been proved to play important roles in technological and device applications \cite{optical1, optical2}.

\medskip
\par
Our calculated tight-binding band structures are consistent with the first-principle calculations at low energy, regarding both the band gap and energy dispersion. For unstrained system, the pair of low-lying valence and conduction bands are mainly governed by the $p_z$ orbitals of  N and B atoms, respectively. This is the signature of a typical bilayer p-n junction where each single layer contributes to either unoccupied or occupied states. Furthermore, the valence and conduction bands are separated by a small direct band gap $\sim 0.2$ eV, featuring a semiconducting characteristic. The presence of  biaxial strain can give rise to structural distortion as well as significant change in atomic interactions. Consequently, the   electronic structure is subjected to a substantial alteration, including changing energy dispersion and band gap as well as inducing band inversion (see Fig. 2b). The band gap can either be enlarged or reduced by applying tensile or compressive strain accordingly. In addition, while tension does not change the orbital distribution of the valence and conduction bands, a suitable compression $\epsilon (\mathbf{d}_s) < -1\%$ can do, leading to band inversion near the M point. It is also noticed that the biaxial strain hardly changes the band gap within the range where the band inversion takes place. Instead, the induced rearrangement of orbital distribution alters significantly the band-edge states (the first panel of Fig. 2b), which can lead to interesting physical phenomena.
\medskip
\par

\begin{figure}[htbp]
\begin{center}
\includegraphics[width=0.9\linewidth]{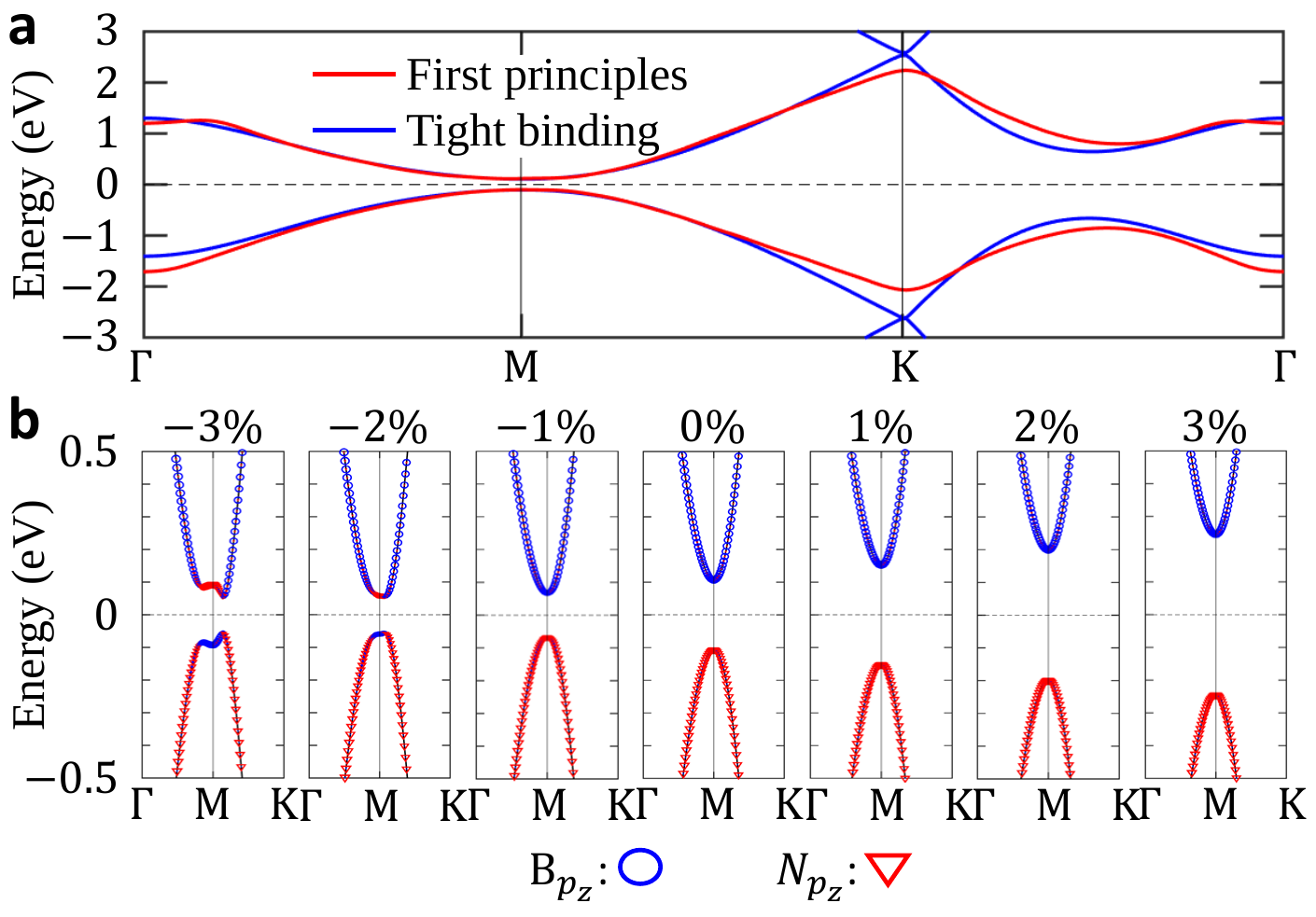}
\end{center}
\caption{(color online) Band structure of C$_3$B/C$_3$N vdW HTS under strain along high-symmetry points. Part (a) shows the band structures of unstrained system. The red lines indicates the first-principles results extracted from Ref. \cite{c3bc3n-pss}, the blue line represents our calculated tight-binding bands. Part (b) illustrates the bandgap modification and evolution of orbital distribution under various biaxial strains. The $p_z$ orbitals of B and N atoms are marked by blue circles and red triangles, respectively.}
\label{Fig2}
\end{figure}

It is worth mentioning that we observe the band inversion, a key signature of a topologically nontrivial material \cite{bandinversion-prb}, in non-topological materials. We also provide below a persuasive explaining why the band inversion is observed, based on its close connection with the lattice properties. This will add an important contribution to the extensive understanding of the causes for band inversion and  the basic inferences when it occurs. It was suggested that such phenomenon in the topological insulators is usually induced by  spin-orbit coupling (SOC) \cite{bandinversion-prl}. In fact, band inversion is not necessarily triggered by SOC, but instead by other factors such as scalar relativistic effects and lattice distortions \cite{bandinversion-aip}. Here, we use the Wilson loop \cite{wilson} to examine the topological properties and confirm that C$_3$B/C$_3$N vdW HTS with stress within $[-3, 3] \%$ are non-topological materials. It is noted that SOC is neglected in our theoretical model.
Our calculations show that the band inversion in C$_3$B/C$_3$N vdW HTS can occur as a result of lattice defect induced by biaxial strain. It was previously reported that C$_3$B/C$_3$N vdW HTS has a strong built-in electrical field which is responsible for the band inversion \cite{c3bc3n-jpcl}. This is also reflected in our TBM where such strong electric field is associated with a significant difference in the on-site energy of B and N atoms.
Interestingly, the band inversion in C$_3$B/C$_3$N vdW HTS exists without the crossing of valence and conduction bands, different from what was proposed in many studies.

\medskip
\par

\begin{figure}[htbp]
\begin{center}
\includegraphics[width=0.9\linewidth]{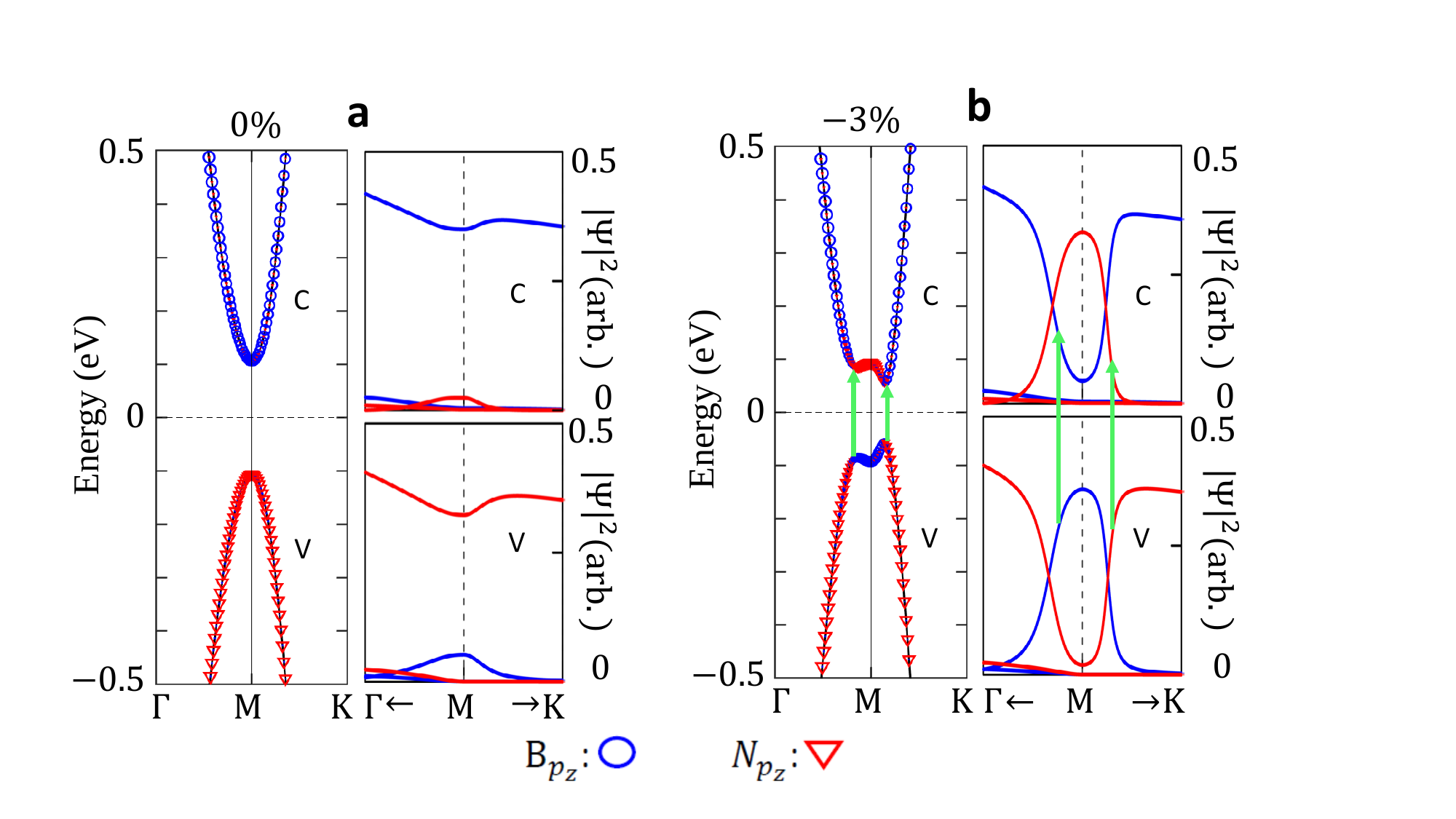}
\end{center}
\caption{(color online) The orbital distribution probability on the low-lying valence and conduction bands for (a) $\epsilon (\mathbf{d}_s) = 0\%$ and (b) $\epsilon (\mathbf{d}_s) = -3\%$. V and C denote the valence and conduction bands, respectively. The vertical green arrows in (b) indicate the significant absorption transition channels between the valence and conduction states. }
\label{Fig3}
\end{figure}
The orbital distribution probability, $|\Psi|^2$, is critical for the understanding of the absorption transition between the valence and conduction states. It is worth mentioning that, only electronic transition between the same type of atomic orbitals exists. Therefore, the available absorption transition channels are decided by the type of orbitals of the occupied and unoccupied states while the intensity of spectral peaks is determined by the amplitude of $|\Psi|^2$. For C$_3$B/C$_3$N vdW HTS, both B and N $p_z$ orbitals play the dominant role in the considered energy range, referring to Figs. 3a and 3b. In the absence of band inversion (see Fig. 3a), the valence band is clearly dominated by $N_{p_z}$ orbital whereas the conduction band is mainly occupied by $B_{p_z}$ orbitals. As a result, no noticeable interband absorption transition is expected. When the band inversion is triggered by a suitable applied strain, the rearrangement of orbital distribution can be clearly seen through $|\Psi|^2$, as shown in Fig. 3b for $\epsilon (\mathbf{d}_s) = -3\%$. For the valence band around the M point where the band inversion takes place, the distribution probability of $N_{p_z}$ is decreased as that of $B_{p_z}$ is increased substantially. The opposite is true for the conduction band. During such a process, both $N_{p_z}$ and $B_{p_z}$ have considerable contribution to the valence and conduction bands. This leads to significant $N_{p_z} \to N_{p_z}$ and $B_{p_z} \to B_{p_z}$ absorption excitation channels, as marked by the vertical green arrows in Fig. 3b.
\medskip
\par

\begin{figure}[htbp]
\begin{center}
\includegraphics[width=0.9\linewidth]{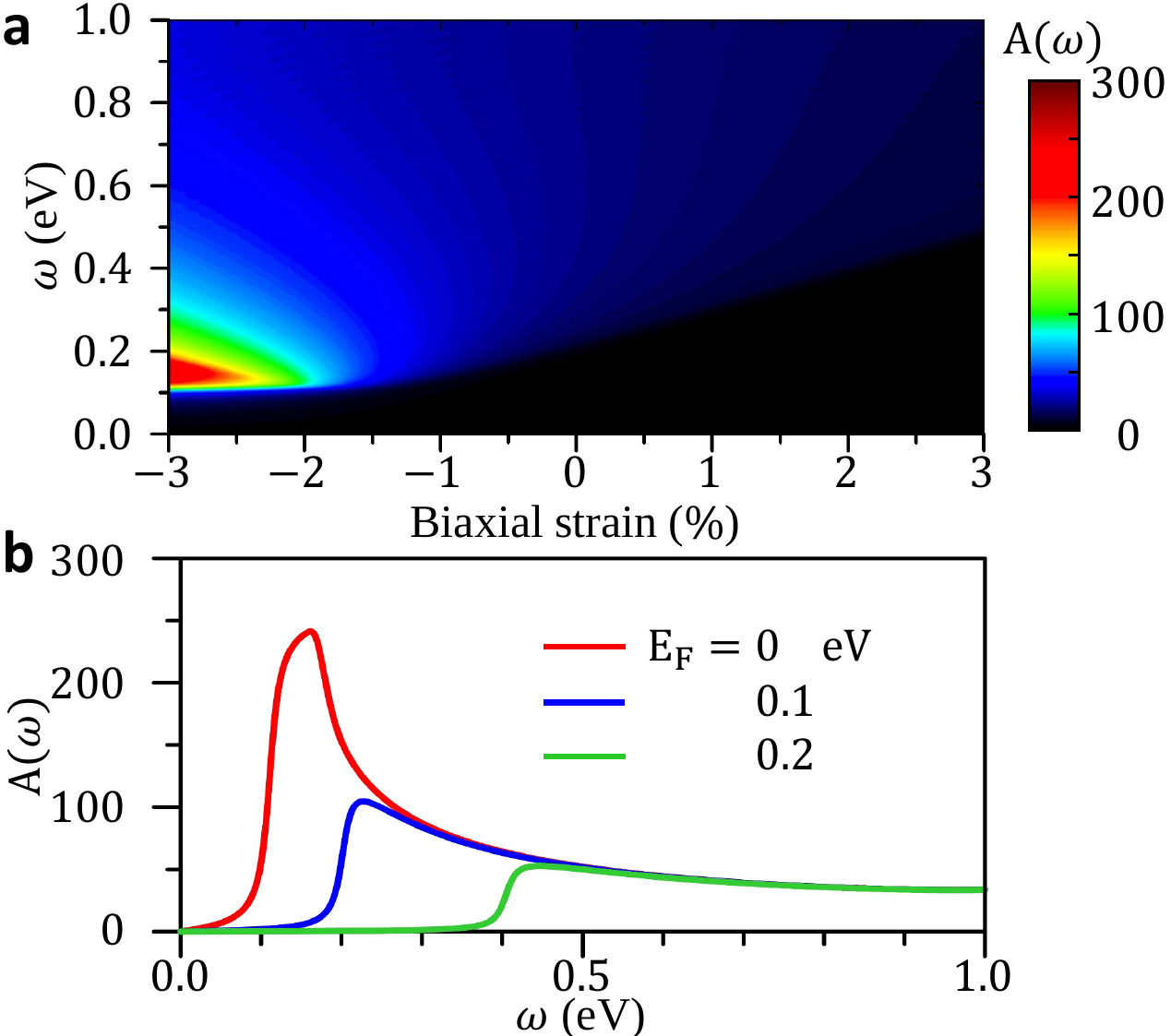}
\end{center}
\caption{(color online) (a) The (strain, frequency)-phase diagram for absorption-optical spectrum. The vertical bar is the color scale for the absorption coefficient, corresponding to the spectral weight. A closer look at the absorption threshold structure for both undoped and doped C$_3$B/C$_3$N vdW HTS is presented in (b). We consider various Fermi energies for the case when the band inversion occurs at strain $\epsilon (\mathbf{d}_s) = -3\%$. }
\label{Fig4}
\end{figure}
For undoped C$_3$B/C$_3$N vdW HTS, the Fermi level lies within the band gap. Thus, the absorption spectrum is dominated by interband transitions and the optical threshold frequency $\omega_{0}$ is equal to the band gap. The strain engineering of electronic structure plays a crucial role in analyzing the optical characteristics under strain. The change of band gap corresponds to the evolution of $\omega_{0}$, referring to the dark area in Fig. 4a. For strain $\epsilon (\mathbf{d}_s) \leq -2\%$ where the band inversion clearly takes place, $\omega_{0}$ reaches the lowest and remains unchanged. This is consistent with the fact that there is no crossing between the conduction and valence bands during the band inversion as we discussed above. Recognizing the influence of strain on $\omega_{0}$ is critical for adjusting the incident light to trigger the photoelectric effect on materials.
We observe special spectral features with very strong weight (red area) for $\epsilon (\mathbf{d}_s) \leq -2\%$, starting from $\omega_{0}$ and extending up to $\omega = 0.2$ eV when $\epsilon (\mathbf{d}_s)$ approaches -3$\%$. This is associated with the increasing density of states (DOS) due to distortion of band-edge states and the change of orbital distribution probability due to the band inversion. Such behavior can enhance apparently the interband transitions threshold regarding both spectral peak intensity and width.
\medskip
\par

Electron doping has been demonstrated efficient in controlling electronic and optical properties of many 2D materials.
Understanding the doping engineering of absorption spectrum of C$_3$B/C$_3$N vdW HTS is crucial for possible designing of electronic and optoelectronic devices.
Let us take a closer look at the special absorption threshold for $\epsilon (\mathbf{d}_s) = -3\%$ for the non-doped system (red curve in Fig. 4b). The peak appears at $\omega \approx 0.1$ eV with obviously asymmetric shape. The peak intensity quickly rises to the highest at $\omega \approx 0.1$ eV, remains high for $\omega \in [0.1-0.2]$ eV, and then gradually declines for higher frequency.
Electron doping can modify significantly the main features of the threshold peak, including the intensity, frequency, and dispersion of peak (Fig. 4b).
Raising $E_F$ can lower substantially the peak intensity while increase its frequency.
Furthermore, the absorption threshold changes gradually from asymmetric peak to shoulder-like structure.
In fact, when $E_F$ is sufficiently high so that the conduction band-edge states become occupied, the interband transition between the valence and conduction band-edge states are forbidden. As a result, the specially high threshold spectrum no longer exists.
In general, higher $E_F$ has influence on wider energy range.
For the considered $E_F$'s, spectrum at higher frequency, $\omega \geq 0.5$ eV, remains unchanged.
\medskip
\par

\begin{figure}[htbp]
\begin{center}
\includegraphics[width=0.9\linewidth]{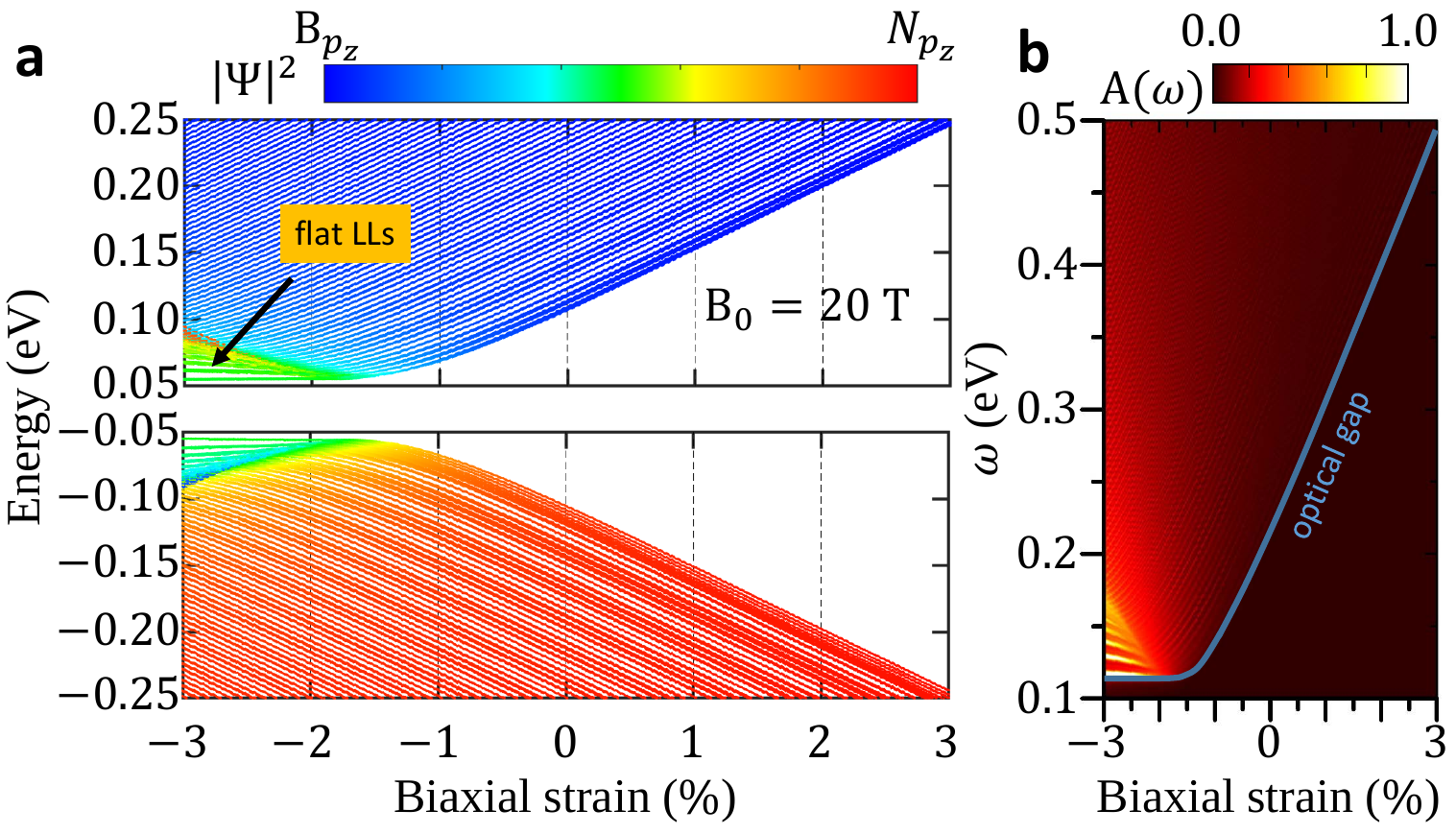}
\end{center}
\caption{(color online) The interplay between an external magnetic field and strain on electronic and optical properties. (a) LL energy spectrum for changing strain strength within the range $\epsilon (\mathbf{d}_s) \in [-3, 3] \%$ at $B_0$ = 20 T. The horizontal bar is the color scale for the density probability of B and N atoms. Part (b) illustrates the influence of strain on the magneto-optical spectrum. The (strain, frequency)-phase diagram for magneto-optical spectrum of non-doped system at T = 0 is presented. The horizontal bar is the color scale for the absorption coefficient, corresponding to the spectral weight.}
\label{Fig5}
\end{figure}
When the system is subject to a uniform perpendicular magnetic field, electronic states are quantized into LLs. For C$_3$B/C$_3$N vdW HTS, the impurities due to atomic doping destroy several geometric symmetries of the system, such as mirror symmetry, inversion symmetry, and rotational symmetry. As a result, the magnetic quantization of the system is rather complex compared with that of graphene monolayer and bilayer. The low-lying LLs are dominated by $B_{p_z}$ and $N_{p_z}$ orbitals instead of $C_{p_z}$ as in pristine graphene systems. For unstrained HTS, the valence and conduction LLs are governed by $N_{p_z}$ and $B_{p_z}$ orbitals, respectively (see Fig. 5a). Generally, the LL features correspond to the zero-field band structure regarding the band gap and energy dispersion. The initial valence and conduction LLs are quantized from the band-edge states, therefore, the band gap separating the occupied and unoccupied LLs is equal to the band gap at B$_0$ = 0.
\medskip
\par

Biaxial strain has a great influence on the magnetic quantization, as demonstrated in Fig. 5a. The evolution of LL band gap and modification of orbital distribution for variation of $\epsilon (\mathbf{d}_s)$ resembles that at B$_0$ = 0. LLs away from zero energy ($E > 0.1$ eV and $E < -0.1$ eV) are governed by either $B_{p_z}$ or $N_{p_z}$ orbitals regardless of $\epsilon (\mathbf{d}_s)$. The effect due to strain on LLs becomes essential in the vicinity of zero energy, corresponding to strain $\epsilon (\mathbf{d}_s) \leq -1\%$ for which the band inversion plays a critical role. Within this region, both $B_{p_z}$ and $N_{p_z}$ orbitals make considerable contribution to the LLs, as a result of the band inversion. This can be understood from the evolution of $|\Psi|^2$ at B$_0$ = 0 as discussed above (see Fig. 3b). There exists a large number of LLs in a narrow energy range for both valence and conduction bands due to the very high DOS. In addition, LLs quantized from the distorted band-edge states are nearly flat. This indicates that there is no crossing between conduction and valence states during the band inversion in our considered strain range. These strain-induced peculiar features of LLs are expected to diversify the absorption spectrum.

\medskip
\par
The strain engineering of the magneto-absorption spectrum is illustrated in Fig. 5b. The blue curve separating the darker and brighter red area presents the strain-dependent optical gap. This is consistent with the progression of LL band gap for the variation of $\epsilon (\mathbf{d}_s)$. That is because the absorption threshold belongs to the optical transition between the initial valence and conduction LLs. The spectral region where the role played by band inversion becomes critical, $\omega < 0.2$ eV and $\epsilon (\mathbf{d}_s) < -2\%$, presents exotic characteristics. There exist several separated nearly-flat white lines, indicating the prominent spectral peaks. These absorption peaks represent the optical transition between the nearly flat valence and conduction LLs. It is worth noting that, each $B_0 = 0$ energy band can be quantized into a large number of LLs. Such a rearrangement of electronic states leads to remarkable changes in absorption peaks. Particularly, the magneto-absorption spectrum presents two-order larger number of peaks with two-order weaker intensity and narrower width compared with that at zero field. These considerable differences between the zero-field- and magneto-absorption spectra signifies the possibility of controlling the optical properties of the strained C$_3$B/C$_3$N vdW HTS by external magnetic field for various device applications.

\medskip
\par

\section{Concluding Remarks}
\label{sec4}

We have presented a thorough theoretical study on the electronic and optical properties of C$_3$B/C$_3$N vdW HTS under a combined effect due to atomic doping, biaxial strain, and external magnetic field. The inclusion of these factors in the calculations was done based on the tight-binding model in conjunction with the Peierls substitution. We found that biaxial strain leads to intriguing electronic and optical characteristics, including the band inversion, modification of band gap and optical gap, distorted band-edge states, and exotic optical-absorption spectrum. The application of an external magnetic field gives rise to more peculiar electronic and optical properties of the strained HTS. We observed the strain-dependent LL energy spectrum with a large number of LLs, nearly flat LLs, and special orbital distribution. Furthermore, proper control of strain can lead to special prominent magneto-absorption spectral peaks at low energy. The main difference between the zero-field- and magneto-absorption spectra lies on the field-induced two-order larger number of peaks with two-order weaker intensity and narrower width. We have provided an indisputable explanation for the extraordinary electronic and optical properties through their close connection with the strain-triggered band inversion and band gap alteration. This work added important information for the understanding  of the engineering of fundamental physical properties of 2D materials by external effects. Our study suggests that C$_3$B/C$_3$N vdW HTS is a promising candidate for next-generation electronic and optoelectronic devices such as flexible strain sensors and photodetectors.
\medskip

\section*{Acknowledgement(s)}
The authors would like to thank the MOST of Taiwan for the support through Grant No. MOST111-2811-M-006-009.
G.G. would like to acknowledge the support from the Air Force Research Laboratory (AFRL) through Grant No. FA9453-21-1-0046.
\medskip

\section*{Competing Interests}
The authors declare no competing interests.
\medskip

\section*{Author contributions}
{\bf P.H.Shih}: Methodology, Formal analysis, Writing-Original Draft, Visualization; {\bf T.N.Do}: Conceptualization, Methodology, Software, Validation, Manuscript revision; {\bf G.Gumbs}: Conceptualization, Validation, Visualization, Supervision.

\end{document}